\documentclass[%
 prl,
 amsmath,amssymb,
 reprint,%
]{revtex4-1}

\usepackage{graphicx}
\usepackage{dcolumn}
\usepackage{bm}

\begin{document}

\title{A Fair Sampling Test for Ekert Protocol}%

\author{Guillaume Adenier}
\email{adenier@rs.noda.tus.ac.jp}
\author{Noboru Watanabe}%
\affiliation{Tokyo University of Science, 2641 Yamazaki, Noda, Chiba 278-8510, Japan}%
\author{Andrei Yu. Khrennikov}
\affiliation{Linnaeus University, Vejdes plats 7, SE-351 95 V\"{a}xj\"{o}, Sweden}%

\begin{abstract}
We propose a local scheme to enhance the security of quantum key distribution in Ekert protocol (E91). Our proposal is a fair sampling test meant to detect an eavesdropping attempt that would use a biased sample to mimic an apparent violation of Bell inequalities. The test is local and non disruptive: it can be unilaterally performed at any time by either Alice or Bob during the production of the key, and together with the Bell inequality test.
\end{abstract}
\maketitle
\section{Introduction}
Ekert protocol \cite{Ekert91,Gisin02,Jaeger07} uses entangled states to guarantee the secrecy of a key distributed to two parties (Alice and Bob). Identical measurements performed on a maximally entangled state yield perfect correlation, which can be used to produce a shared key; while the secrecy of the key can be guaranteed by the violation of Bell inequalities measured for non-identical measurements. An unconditional violation of Bell inequalities would guarantee that no local (hidden) variables exist that an eavesdropper (Eve) could exploit. It would mean unconditional privacy: Eve could have full control of the detectors and the source, more advanced theory and technology, it would still be secure \cite{Gisin02}.

 However, practical implementations of Ekert protocol have to be performed with photons, because a key distribution protocol is useful only if Alice and Bob can be separated by macroscopic distances \cite{Scarani09}. Photons are also restricted by the use of polarizing beam splitter to a wavelength domain for which standard photon counters have a poor detection efficiency \cite{LoZhao,Stucki}. It means that a rather heavy postselection is required: Alice and Bob must discard all measurements for which either of them failed to register a click at all \cite{Gisin02}. The trouble is that local hidden-variable models that exploit this weakness can reproduce exactly the predictions of Quantum Mechanics \cite{Pearle}, as soon as the detection efficiency is lower than $83\%$ \cite{GargMermin,Eberhard,Larsson1,Gisin99}. In the context of experiments on the foundations of Quantum Mechanics, the assumption of Fair Sampling is usually considered reasonable to support a violation of Bell inequalities, with the idea that Nature is not conspiratory. In Quantum Key Distribution however, Eve is expected to conspire \cite{Ekert09}. Alice and Bob should therefore assume that their sample is biased by Eve. Failure to acknowledge this weakness would leave all freedom to Eve to exploit it with a biased sample attack: the statistics on the detected sample would then only have the appearance of secrecy. This weakness should not be underestimated given that a successful quantum hacking has already been successfully implemented experimentally by means of a time-shifting attack \cite{Lo08}.

 Naturally, this issue becomes critical if Eve manufactured the detectors, which means that Alice and Bob should thoroughly check that their detectors are functioning according to specifications \cite{Gisin02}. However, we will argue here that even if the detectors owned by Alice and Bob are genuine photomultipliers or avalanche photodiodes, Eve could still in principle force a biased sampling on these detectors by exploiting the thresholds of these detectors. Eve would only need to control the source and know the detectors well enough to exploit their thresholds, but she would not need to actually control them. We will thus propose below a fair sampling test to prevent such a biased-sample attack on threshold detectors.

\begin{figure}
\center
\includegraphics[width=8cm]{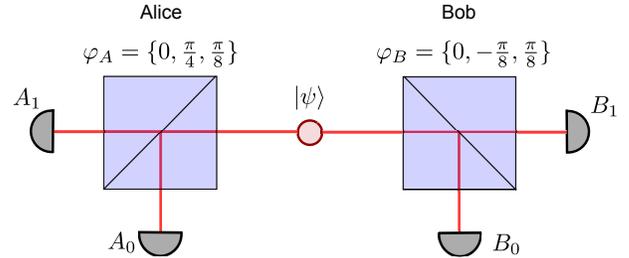}
\caption{\label{fig:epr_setup10}Standard Ekert protocol. Alice and Bob randomly switch their measurement settings. Pairs associated with identical measurement settings ($\varphi_A=\varphi_B$) are used to produce a correlated key, while those associated to non-identical measurement settings ($\varphi_A\neq\varphi_B$) are used to check the violation of Bell inequality (and the security of the key).}
\end{figure}
\section{A biased-sample attack on Ekert protocol}
The motivation and concern for the possibility of a biased-sample attack is that avalanche photodiodes and photomultipliers are fundamentally threshold detectors. At the input, the energy must be higher than the band gap to trigger an avalanche or a photoelectron; while at the output, the current must be higher than a discriminator value to be counted as a click \cite{Knoll}. This combined threshold could be exploited by Eve to obtain an apparent violation of Bell inequalities on the detected sample \cite{Adenier09}.

 We will assume throughout this paper that the source is controlled by Eve, that she can produce pulses that split classically according to Malus law in polarizing beamsplitters, and that these pulses are sensitive to the threshold in Alice's and Bob's detector. How Eve will effectively produce such pulses is left to her, but it should be stressed that if each pulse contains at most one particle then the biased sampling described here would be ineffective, because the energy seen at a detector would always be the same regardless of the measurement settings. Eve could for instance produce pulses with several photons of lower frequencies, possibly using non-linearities in threshold detectors \cite{Resch01}.

We will consider here simple models of threshold detectors: ideal threshold detectors, which produce a click with certainty if the energy $E$ of the absorbed pulse is greater than a threshold $\Phi$; and linear threshold detector, which produce a click with a probability increasing linearly with the energy above a threshold $\Phi$ (possibly with a saturation value after which the probability no longer increases).

The simplest way for Eve to obtain an apparent violation of Bell inequalities reproducing exactly the predictions of Quantum Mechanics on the detected sample is to aim at reproducing the asymmetrical detection pattern of a Larsson-Gisin model \cite{Larsson3,Gisin99}. Those models are \emph{ad hoc}, but it is in fact relatively straightforward for Eve to obtain these patterns with classical pulses and threshold detectors.

For this purpose, Eve sends pairs of correlated pulses with energy $E_0$ and polarization $\lambda$, where $\lambda$ is a random variable uniformly distributed on the interval $[-\pi/2,\pi/2]$. Then she just needs to make sure that on one side (say, Alice) the detectors are ideal threshold detectors with $E_0=2\Phi$, while the other side (Bob) the detectors are linear threshold detectors, also with $E_0=2\Phi$. With the condition $E_0=2\Phi$, Malus' Law is cut by the bottom precisely at the intersection of the two channels (at $|\varphi-\lambda|=\pi/4$). Consequently, Alice's always records a click in exactly one channel: channel $1$ if $|\varphi_A-\lambda|<\pi/4$, channel $0$ otherwise; whereas on Bob's side the probability to get a click in the channel $1$ varies with $\cos 2(\varphi_B-\lambda)$ when $|\varphi_B-\lambda|<\pi/4$, and with $\sin 2(\varphi_B-\lambda)$ when $|\varphi_B-\lambda|>\pi/4$ in channel $0$. The crucial feature of the resulting detection pattern is that the probability to obtain a click in either channel on Bob's side depends explicitly on $\lambda$: it is maximum for $|\varphi_B-\lambda|=0$, and decreases down to zero for $|\varphi_B-\lambda|=\pi/4$. The sampling is thus unfair, or biased, and leads to an apparent violation of Bell inequalities on the detected sample \cite{Larsson3,Gisin99,Adenier09}.

An eavesdropping strategy would therefore consist in replacing the source of entangled photons with a classical source of pulsed pairs correlated in polarization and designed to meet condition $E_0=2\Phi$. If Eve aims at reproducing the full correlation function as predicted by Quantum Mechanics, she would have to make sure that at one station (say, Alice) the threshold detectors react ideally to the pulses, whereas at the other station (Bob) the threshold detectors react linearly. However, if Alice and Bob are only measuring a few points of the correlation function (those giving maximum violation of Bell inequality), as is done in Ekert protocol, Eve can lift this constraint and work with identical threshold detectors on both sides (either linear or ideal). Alice and Bob would then observe a maximal violation of Bell inequalities on the subset of detected pairs, and would thus wrongly believe that their key is secure while in fact Eve's knowledge would in principle be maximum.
\section{Countermeasure: a fair sampling test}
In order to prevent Eve from using this attack, the obvious solution consists in increasing the efficiency to reach 83\%. However, this proves difficult with threshold detectors. Decreasing the band gap threshold---or increasing the operating temperature---does increase the efficiency of the detectors, but only at the cost of higher dark count rates. Unless special detectors operating near absolute zero temperature are used, such as Transition-Edge Sensors (which are too cumbersome and slow to be practical solution to QKD), this can be considered a general rule that applies to any detectors, and fundamentally limits their efficiencies.

Another suggestion is to artificially complete the detected sample by randomly assigning 0 or 1 to non-detected pulses \cite{Ma}, so that the required efficiency to produce a useful key is lowered to $50\%$. However, we would like to argue that the drawback of this method is that introducing some random results would be bound to decrease the violation of Bell inequality measured on the completed sample, thus preventing a security check of the key unless one does so on the uncompleted sample (which would again reintroduce the $83\%$ efficiency bound).

\begin{figure}
\center
\includegraphics[width=8cm]{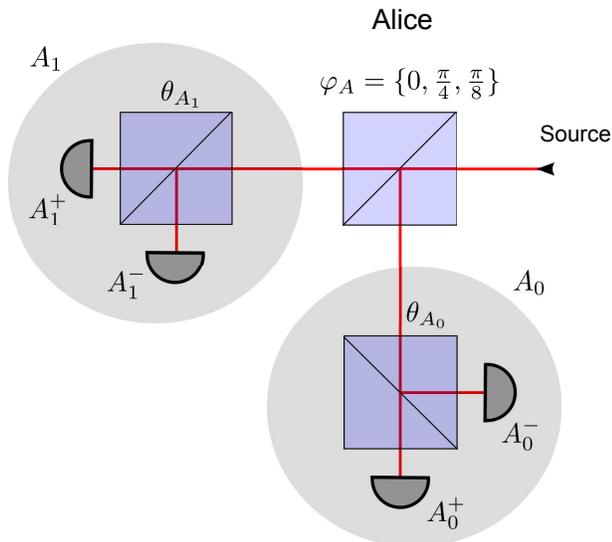}
\caption{\label{fig:fstest10}Fair Sampling test on Alice's side. The detector in channel $1$ is replaced by a polarimeter $A_1$ with two detectors $A_1^+$ and $A_1^-$ having the same efficiency $\eta$. The detector in channel $0$ is replaced by a polarimeter $A_0$ with two detectors $A_0^+$ and $A_0^-$, also with efficiency $\eta$. Ekert protocol is thus unaltered by our test: polarimeter $A_1$ is equivalent to the detector in channel $1$ in Fig.~\ref{fig:epr_setup10}, with the same efficiency $\eta$, and polarimeter $A_0$ is equivalent to the detector in channel $0$ with efficiency $\eta$. Similar results would be obtained for polarimeter $A_0$, and for Bob's polarimeters.}
\end{figure}

Our proposal consists in testing the fairness of the sample by analyzing the output channels of the polarizing beamsplitters, instead of simply feeding detectors with them. We keep the standard design of Ekert protocol, with two polarizing beamsplitters on each side (Alice and Bob) projecting the incoming pulses on random bases $\varphi_A$ and $\varphi_B$, as depicted on Fig.~\ref{fig:epr_setup10}, but we replace each detectors by a \emph{polarimeter} \cite{AspectTh}: a polarizing beamsplitter followed by a detector at each output. Consider Alice's side (see Fig. \ref{fig:fstest10}). We label the polarimeter in channel $1$ as $A_1$, the orientation of its polarizing beam-splitter as $\theta_{A_1}$, and the detectors in the transmitted and reflected output as $A_1^+$ and $A_1^-$ respectively. Similarly, the polarimeter in channel $0$ is labeled $A_0$, the orientation of its polarizing beam-splitter $\theta_{A_0}$, and the detectors in the transmitted and reflected output are $A_0^+$ and $A_0^-$ respectively. Bob would proceed similarly with two polarimeters labeled $B_1$ and $B_0$.

In case of a genuine source of entangled photons, nothing is changed for Ekert protocol, as long as all the detectors have the same efficiency $\eta$. Each polarimeter can then be considered as one detector with quantum efficiency $\eta$. The polarimeter $A_1$ can be seen as one single detector in channel 1, in which the orientation $\theta_{A_1}$ has no influence on the result: a photon exiting the polarizing beam-splitter $\varphi_A$ through channel $1$ will be detected in either output channel of polarimeter $A_1$ with a probability $\eta$. Similarly, polarimeter $A_0$ can be seen as one single detector in channel $0$, where the orientation $\theta_{A_0}$ plays no role whatsoever, and the same goes for Bob's setup. The production of the key and the verification of the violation of Bell inequalities is thus unaltered by our fair sampling test setup in case of a genuine source of entangled photons, because the additional measurement settings $\theta_{A_1}$, $\theta_{A_0}$, $\theta_{B_1}$ and $\theta_{B_1}$ controlled by Alice and Bob have no influence on the measurement results in case of a genuine source of entangled photons.

However, they have a strong influence in the case of a biased-sample attack by Eve. Let us consider the simpler case of ideal threshold detectors mentioned above. By Malus law, the energy of the pulse reaching Alice's $A_1^+$ detector is
\begin{equation}\label{E:EA1}
    E_{A_1^+}=E_0 \cos^2(\varphi_A-\theta_{A_1}) \cos^2(\varphi_A-\lambda).
\end{equation}

Starting from a uniform distribution of the polarization $\lambda$ of pulses on the circle, we write that $|P_\lambda d_\lambda|=|P_{A_1^+}(E_{A_1^+}) dE_{A_1^+}|$, so that the probability to get an energy between $E_{A_1^+}$ and $E_{A_1^+}+dE_{A_1^+}$ in the $A^+$ channel is given by
\begin{equation}\label{PA1E}
    P_{A_1^+}(E_{A_1^+})dE_{A_1^+}=\frac{dE_{A_1^+}}{\pi\sqrt{(E_{\rm max}-E_{A_1^+})E_{A_1^+}}}
\end{equation}
where $E_{\rm max}=E_0\cos^2(\varphi_A-\theta_{A_1})$ is the maximum energy reaching the detector (by Malus' law).

The probability to obtain a click in an ideal threshold detector placed at the transmitted output ($+$) of polarimeter $A_1$ is then simply the integral of this density distribution over the energy reaching the detector, from the threshold $\Phi$ to $E_{\rm max}$:
\begin{align}\label{PA1p}
    P_{A_1^+}(E_0,\Phi)&=\int_\Phi^{E_{\rm max}}\frac{dE_{A_1^+}}{\pi\sqrt{(E_{\rm max}-E_{A_1^+})E_{A_1^+}}}\\
    &=\frac{2}{\pi}\arccos\sqrt{\frac{\Phi}{E_0\cos^2(\varphi_A-\theta_{A_1})}}.
\end{align}
 Similarly the probability to obtain a click in an ideal threshold detector positioned at the reflected output ($-$) of polarimeter $A_1$ is
\begin{equation}\label{PA1m}
    P_{A_1^-}(E_0,\Phi)=\frac{2}{\pi}\arccos\sqrt{\frac{\Phi}{E_0\sin^2(\varphi_A-\theta_{A_1})}}.
\end{equation}

\begin{figure}
\center
\includegraphics[width=9cm]{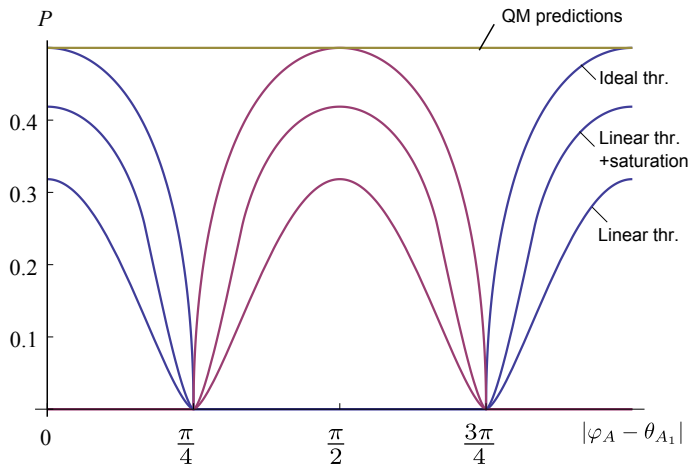}
\caption{Analytical Results in case of biased-sample attack on threshold detectors in Alice's polarimeter $A_1$, with $E_0=2\Phi$. The probability to get a click in $A_1$ depends on $|\varphi_A-\theta_{A_1}|$. From $0$ to $\frac{\pi}{4}$ and from $\frac{3\pi}{4}$ to $\pi$, only detector $A_1^+$ can click, whereas from $\frac{\pi}{4}$ to $\frac{3\pi}{4}$ only detector $A_1^-$ can click. By contrast, in case of a genuine source of entangled photons, the probability to get a click in these channels is governed by Malus law's $\cos^2(\varphi_A-\theta_{A_1})$ and $\sin^2(\varphi_A-\theta_{A_1})$ (not shown here), and the probability to get a click in either channel therefore adds up to a constant independent of $|\varphi_A-\theta_{A_1}|$.}
\label{fig:fulltestresult}
\end{figure}

In the case of linear threshold detectors, the analytical results are more complicated since the probability to get a click for an energy $E+dE$ is not always equal to 1, but the principle of calculation remains the same: integrate the product of the probability density distribution by the probability of obtaining a click for a given energy. The analytical results for linear threshold are qualitatively similar to that of ideal threshold detectors. The results in the case $E_0=2\Phi$---which is leading to a violation of Bell inequalities exactly reproducing the predictions of Quantum Mechanics---are displayed in Fig.~\ref{fig:fulltestresult}: the probability to get a click in polarimeter $A_1$ depends on $|\varphi_A-\theta_{A_1}|$. It is maximum for $|\varphi_A-\theta_{A_1}|=0+k\pi/2$, and reaches zero for $|\varphi_A-\theta_{A_1}|=\pi/4+k\pi/2$. Similar results would be obtained for polarimeter $A_0$, and for Bob's polarimeters.

This fair sampling test can be implemented very simply on Alice's side by fixing $\theta_{A_1}=\theta_{A_0}=0$. The random switching in Ekert protocol (Fig.~\ref{fig:epr_setup10} and Fig.~\ref{fig:fstest10}) ensures that the points at $0$ and $\pi/4$ are both scanned automatically. Any significant difference in the number of single counts recorded when $\varphi_A=0$ and $\varphi_A=\pi/4$ would betray Eve's attempt to bias the sample through a biased-sample attack on the threshold detectors. Similarly, Bob would chose $\theta_{B_1}=\theta_{B_0}=\pi/8$, and compare the number of singles when $\varphi_B=-\pi/8$ and $\varphi_B=\pi/8$.
\section{Conclusion}
 Our fair sampling test can be implemented during the production of the key and together with the violation of Bell inequality check, so that it seems hard to bypass it without reducing the visibility of the correlation. For instance, increasing the energy of the pulses with respect to the threshold would tend to reduce the dip in the fair sampling test, but it would give rise to double counts and reduce the visibility of the correlation at the same time (weaker violation of Bell inequalities). The combination of a Bell inequality test with a monitoring of the double counts and our local fair sampling test therefore constitutes a solid scheme against eavesdropping a E91 protocol using a biased-sample attack. It should also be noted that the use of four detectors on each side can serve other purposes, like shielding Alice and Bob from a time-shift attack \cite{Lo10}. In principle, similar fair sampling tests could be implemented in other QKD protocol, by replacing passive detectors in each channel by a device with the same efficiency that would analyze further whichever degree of freedom is used to encode the key, instead of simply feeding detectors with it.
\section{acknowledgements}
We are grateful to Hoi-Kwong Lo, Jan-{\AA}ke Larsson, Takashi Matsuoka and Masanori Ohya for useful discussions on Quantum Key Distribution.

\end{document}